\documentclass[conference]{IEEEtran}

\usepackage{graphicx}
\usepackage{subfig}
\usepackage{listings}
\usepackage{color}
\usepackage{algorithm2e}
\usepackage{amsmath}
\usepackage{todonotes}
\usepackage{balance}
\usepackage{url}
\usepackage{enumitem}

\DeclareGraphicsExtensions{.pdf,.png}

\begin{document}
\title{A Linear Algebra Approach to Fast DNA Mixture Analysis Using GPUs}

\author{
	\IEEEauthorblockN{Siddharth Samsi, Brian Helfer, Jeremy Kepner, Albert Reuther and  Darrell O. Ricke
	\IEEEauthorblockA{MIT Lincoln Laboratory \\ 
          Lexington, MA}
	}
}

\maketitle
\begin{abstract} Analysis of DNA samples is an important step in forensics, and the speed of analysis can impact investigations. Comparison of DNA sequences is based on the analysis of short tandem repeats (STRs), which are short DNA sequences of 2-5 base pairs. Current forensics approaches use 20 STR loci for analysis. The use of single nucleotide polymorphisms (SNPs) has utility for analysis of complex DNA mixtures.  The use of tens of thousands of SNPs loci for analysis poses significant computational challenges because the forensic analysis scales by the product of the loci count and number of DNA samples to be analyzed. In this paper, we discuss the implementation of a DNA sequence comparison algorithm by re-casting the algorithm in terms of linear algebra primitives. By developing an overloaded matrix multiplication approach to DNA comparisons, we can leverage advances in GPU hardware and algoithms for Dense Generalized Matrix-Multiply (DGEMM) to speed up DNA sample comparisons. We show that it is possible to compare 2048 unknown DNA samples with 20 million known samples in under 6 seconds using a NVIDIA K80 GPU.
\end{abstract}

\IEEEpeerreviewmaketitle

\section{Introduction}
\let\thefootnote\relax\footnotetext{DISTRIBUTION STATEMENT A. Approved for public release: distribution unlimited. This material is based upon work supported by the Assistant Secretary of Defense for Research and Engineering under Air Force Contract No. FA8721-05-C-0002 and/or FA8702-15-D-0001. Any opinions, findings, conclusions or recommendations expressed in this material are those of the author(s) and do not necessarily reflect the views of the Assistant Secretary of Defense for Research and Engineering.} DNA forensics is the branch of forensic science that focuses on the use of genetic material in criminal investigations~\cite{dnaforensics}. Short tandem repeats (STRs) are stretches of DNA containing short repeat units of of neucleotides that are used in forensic DNA and human identity testing~\cite{butler2007short}. DNA forensics currently uses STRs for 20 chromosomal locations, referred to as the Combined DNA Index System (CODIS) loci. Comparing STR profiles between samples and individuals is the current standard for justice systems. Samples with more than one DNA contributor are difficult or impossible to analyze using only STR profiles. Profiling single nucleotide polymorphisms (SNPs) has advantages over STRs for comparisons with mixture samples~\cite{isaacson2015}. In the United States, the Federal Bureau of Investigation (FBI) has a database of over 16 million profiles in the National DNA Index System (NDIS). Comparing a large number of DNA profiles with this large dataset of known reference DNA profiles is currently a computationally expensive process and is typically done in a large datacenter. The FastID~\cite{darrell2017} method was developed to enable rapid searching of forensic panels with large numbers of loci and runs on x86 processors. In this paper we cast the FastID method as a dense matrix multiplication operation and use graphics processing units (GPUs) to enable very fast comparisons between profiles of individuals to individuals, individuals to mixtures, and mixtures to mixtures.

The paper is organized as follows: Section ~\ref{sec:dna_analysis} describes the process of DNA analysis for forensics applications. Section~\ref{sec:algo} gives an overview of the FastID method for DNA mixture comparisons, and in Section ~\ref{sec:matmul} we describe the problem as a dense matrix multiplication algorithm. In Section~\ref{sec:gpu_impl}, details of the GPU implementation of the FastID algorithm and optimizations are described. Finally, in Section ~\ref{sec:results} we present the results of our approach when used to analyze large DNA datasets and we summarize in Section~\ref{sec:summary}.

\section{DNA Mixture Comparison}
\label{sec:dna_analysis}
DNA is composed of a series of molecules called nucleotides and are encoded as A, C, G and T corresponding to the four types of nucleotides. An allele is a variant of a gene that is located at a specific position on a specific chromosome. A single nucleotide polymorphism (SNP) is a genetic variation between individuals and represents a difference in a single nucleotide in a DNA sample. On average there are  10 million SNPs in the human genome~\cite{snpref}. SNPs can act as biological markers of disease and can be used for identifying inheritance within families. In the context of DNA forensics, comparing SNPs in DNA samples can help identify individuals or relatives.

A SNP typically has a major allele that is most common in a population of people and a minor allele with a lower allele frequency than the major allele. Most SNPs have typically only two alleles but more alleles are possible. Let \texttt{M} represent a major allele and \texttt{m} respresent a minor allele. With two alleles for a SNP, there are four possibilities for the SNP for an individual: \texttt{MM}, \texttt{Mm}, \texttt{mM}, and \texttt{mm}. To compare a set of SNPs of size N between two individuals, \texttt{2N} comparisons are needed to compare all alleles.

\begin{figure*}[ht]
  \centering
  \includegraphics[width=35pc]{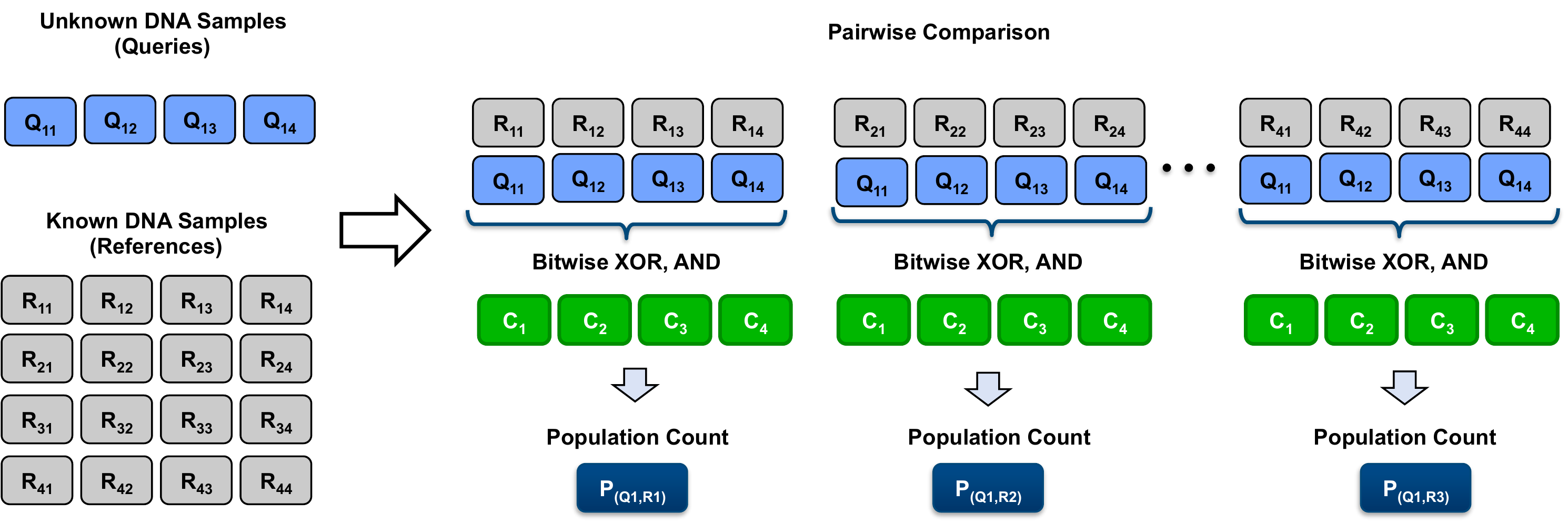}
  \caption{Algorithm for DNA mixture analysis: An unknown DNA sample is compared against each known DNA sequence.}
  \label{fig:algorithm}
\end{figure*}

\subsection{Algorithm for SNP comparison}
\label{sec:algo}
The FastID DNA mixture comparison algorithm used in this paper was first developed by Ricke~\cite{darrell2017}. This algorithm can be used to compare DNA samples from individuals as well as mixtures of samples. The algorithm identifies the similarity between two samples by first performing a bitwise exclusive-OR (XOR) operation between the reference (known) DNA sample and the query (unknown) DNA sample as shown in Figure~\ref{fig:algorithm}. The next step is to perform a bitwise AND operation between this result and the reference sample. Finally, a count of the number of set bits in the result of the AND operation gives a measure of the similarity between the known and unknown DNA sample. In practice, DNA samples can be compared by mapping the string SNP alleles to binary representations and comparing the profiles directly with the computer hardware XOR instruction.  The 1-bits in the result represent all positions where there is a difference in the minor alleles between the two individuals.  The computer hardware population count (POPCOUNT) instruction can then be used to sum the 1-bits in the result to identify all of the minor allele differences between the two profiles. To compare an individual sample to a mixture, a logical AND operation is performed between the XOR results and the individual profile to only consider the minor alleles of the individual.

Let $R_i$ be the reference DNA sample and $Q_j$ be the unknown DNA sample. The similarity between the two samples as quantified by the population count $P_{ij}$ is given by
\begin{equation}
	P_{ij} = POPCOUNT( AND( XOR( R_{i}, Q_{j} ), R_{i} ) )
	\label{eq:popc}
\end{equation}

In the implementation of the FastID algorithm, the DNA samples are first converted from alleles to an array of unsigned integers. A DNA sample with 512 SNPs can be mapped to 16 unsigned 32-bit integer numbers. A 512 SNP DNA sample is thus represented by a length 16 array of unsigned integers. For example, let's consider a DNA sample with 32 SNPs: $0x06001440$. The binary representation of this SNP is $0000 0110 0000 0000 0001 0100 0100 0000$ and the 32-bit unsigned integer decimal equivalent of this is $100668480$. This procedure is used to convert all known and unknown DNA samples into arrays of 32-bit unsigned integers. The algorithm proceeds by performing the operation in Equation~\ref{eq:popc} for each integer in the arrays representing the known and unknown DNA samples. The length of the array depends on the number of SNPs used in the comparison and will be denoted by $N_W$ in the rest of the paper. The algorithm for comparing a single unknown DNA mixture of legnth $N_W$ with a known sample of the same length is shown in Algorithm~\ref{alg:single}. This algorithm can be viewed as an overloaded dot-product of two vectors of length $N_W$ where the multiplication operation is replaced by sequence of logical XOR and AND operations followed by the population count (POPCOUNT) operation.

\begin{algorithm}[h]
  \KwData{Known DNA sample $R$ of length $N_W$}
  \KwData{Unknown DNA mixture $Q$ of length $N_W$}
  \KwResult{Population count $P$}
  initialization\;
  \For{ i = 0 to $N_W - 1$}{
    $A = XOR ( R[i], Q[i] )$ \\
    $B = AND ( A, R[i] ) $ \\
    $Popcount[i] = Popcount[i] + POPC(B) $ \\
  }
  \vspace{.25cm}
  \caption{The core implementation the SNP comparison algorithm: A single known DNA sample $R$ of length $N_W$ is compared with an unknown mixture $Q$ of the same length.}
  \label{alg:single}
\end{algorithm}

In practice, law enforcement agencies such as the Federal Bureau of Investigation (FBI) have millions of known DNA profiles and a correspondingly large number of unknown samples that need identification. Let $N_R$ be the number of known DNA samples and $N_Q$ be the number of unknown samples, each of length $N_W$ as described previously. The algorithm in Algorithm~\ref{alg:single} can now be re-written as shown in Algorithm~\ref{alg:naive2}. The operation in Equation~\ref{eq:popc} must now be performed $N_R * N_Q * N_W$ times.

\begin{algorithm}
  \KwData{$N_R$ known DNA samples of length $N_W$}
  \KwData{$N_Q$ unknown DNA samples/mixtures length $N_W$}
  \KwResult{$P_{QR}$ Population counts}
  initialization\;
  \For {i = 0 to $N_Q - 1 $} {
  	\For {j = 0 to $N_R - 1$} {
      \For{ k = 0 to $N_W - 1$}
      { $A = XOR ( R_i[k], Q_j[k] )$ \\
        $B = AND ( A, R_i[k] ) $ \\
        $Popcount[i,j] = Popcount[i,j] + POPC(B) $ \\
      } } }
      \vspace{.25cm}
  \caption{A na\"ive implementation the SNP comparison algorithm for $N_Q$ individuals and $N_R$ mixtures.}
  \label{alg:naive2}
\end{algorithm}

\begin{figure*}[ht]
  \centering
  \includegraphics[width=35pc]{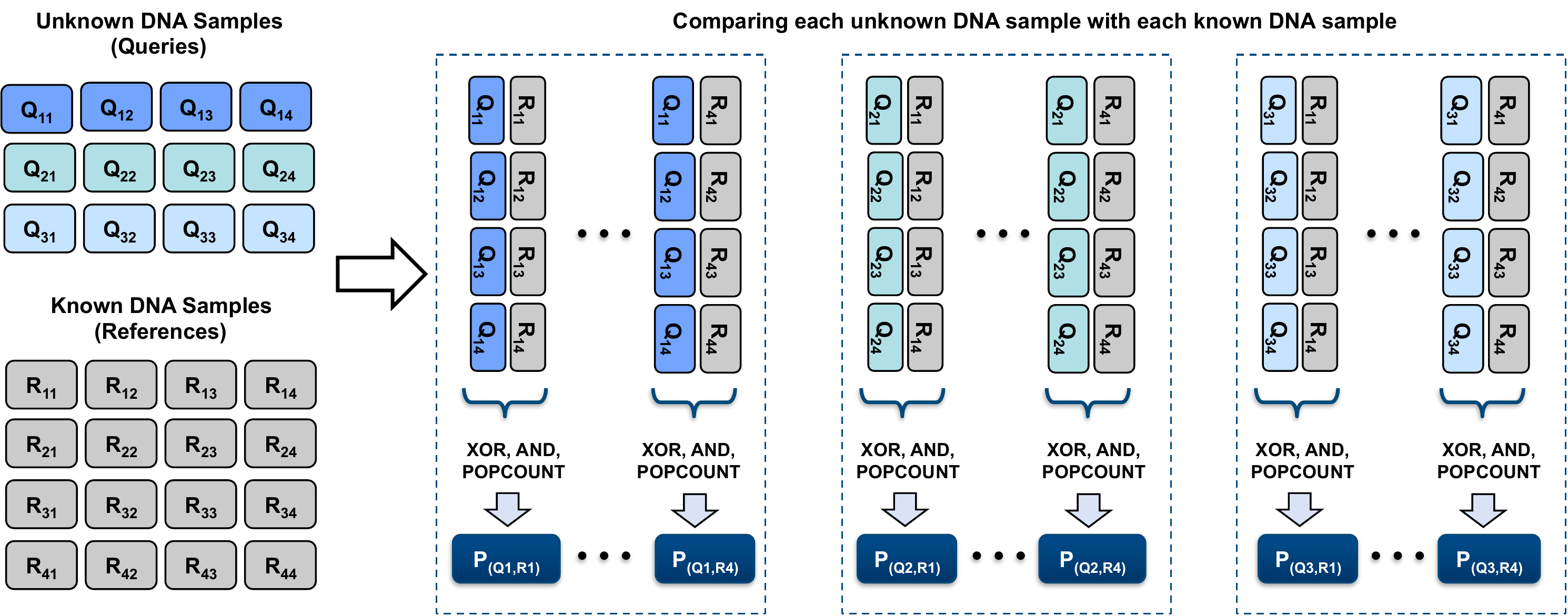}
  \caption{Algorithm for DNA mixture analysis: Each unknown DNA sample is compared against each known DNA sequence.}
  \label{fig:algorithm2}
\end{figure*}

\begin{figure}[hb]
  \centering
  \includegraphics[width=20pc]{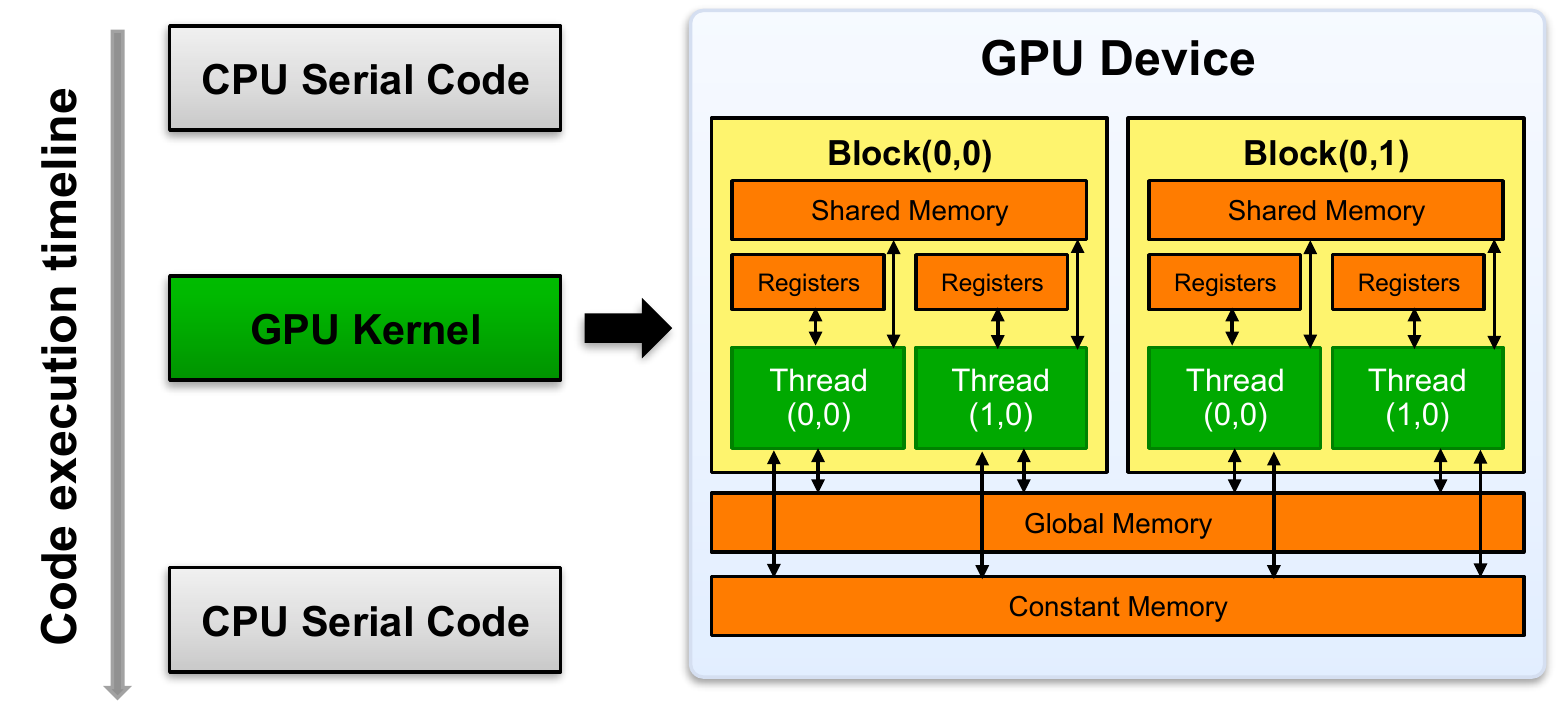}
  \caption{CUDA program execution and GPU memory architecture, after~\cite{kirk2013}.}
  \label{fig:gpu_mem_arch}
\end{figure}

\subsection{DNA Comparison as Matrix Multiplication}
\label{sec:matmul}
Given $N_R$ known DNA samples of length $N_W$ and $N_Q$ unknown DNA mixtures of length $N_W$, the goal is to compare every unknown sample with every known sample. In this case, we can now view this procedure as an overloaded dot product of $N_Q$ vectors representing unknown samples with each of the $N_R$ known samples as shown in Figure~\ref{fig:algorithm2}. We cast the proposed algorithm as a dense matrix multiplication operation by organizing the input data into two matrices of size $N_R$ x $N_W$ and $N_W$ x $N_Q$ representing the known and unknown samples, respectively. Thus, the population counts for a given set of DNA samples can be represented by the overloaded matrix multiplication operation $C = AB$, where A is of dimension $N_R$ x $N_W$, B is of dimension $N_W$ x $N_Q$ and C is of dimension  $N_R$ x $N_Q$. The matrix multiplication is overloaded as shown in Equation~\ref{eq:popc}, where the multiply operation in the matrix multiplication algorithm is replaced by a logical XOR and AND operations followed by the POPCOUNT operation. 

\section{DGEMM on GPU for Mixture Analysis}
\label{sec:gpu_impl}

\subsection{GPU Architecture}
\label{sec:gpuarch}
The algorithm described in this paper was developed on the NVIDIA TESLA K80 GPU and will be referred to as K80 in the remainder of the paper. The K80 consists of 4992 NVIDIA CUDA cores in a dual-GPU design with an aggregate 24GB GB of GDDR5 memory~\cite{k80}. The processing described in this paper used a single GPU in the K80.

Figure ~\ref{fig:gpu_mem_arch} shows the execution of a program written using the NVIDIA CUDA programming platform and language and the memory hierarchy of NVIDIA GPUs. The serial code runs on the CPU and the parallel section of the code, implemented using the CUDA library is launched on the GPU kernel. The CUDA programming model enables programmers to run fine-grained parallel code on the GPU on a large number of threads~\cite{sidspm}. Threads are organized into grid blocks as shown in Figure~\ref{fig:gpu_mem_arch}. A block is a group of threads that runs on a single multiprocessor where they have access to 64KB of shared memory on the K80. A collection of threads that run concurrently on the GPU is called a warp. For detailed descriptions of the execution of a CUDA program, the reader is referred to Kirk \& Wu~\cite{kirk2013}. The GPU also has several types of memory available to each individual thread: global, shared and constant memory. Constant memory is read-only for the threads whereas the global and shared memories can be written to and read by the threads. The amount of shared and constant memory on the GPU is significantly smaller than the global memory but accesses to the shared and constant memory are much faster than global memory. The optimization of CUDA programs involves the management of data transfers to the GPU, data layout in device memory and the maximization of computation to global memory transfer ratio. These optimizations are discussed in Section~\ref{sec:gpu_matmul}.

\subsection{Optimizing overloaded matrix multiplication on GPU}
\label{sec:gpu_matmul}
\begin{figure}[!ht]
  \centering
  \includegraphics[width=20pc]{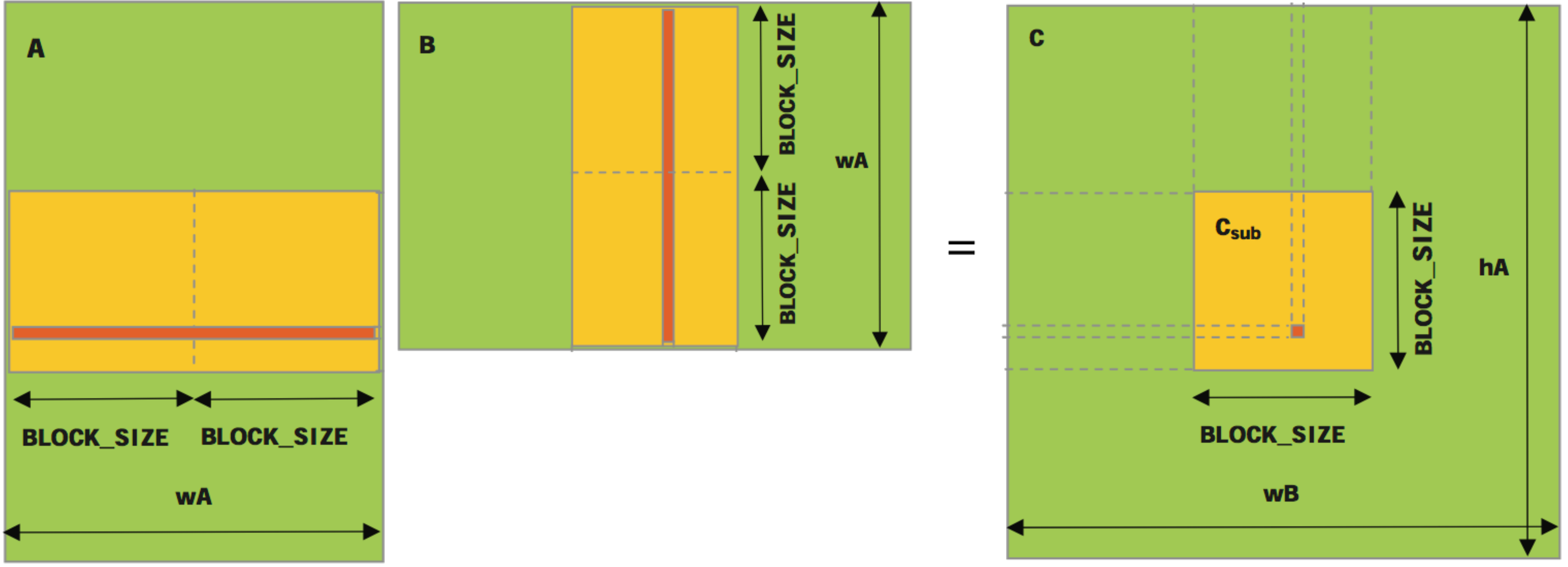}
  \caption{Blocked matrix multiplication: Each thread computes one element of the output matrix~\cite{nvidia}.}
  \label{fig:block_matmul}
\end{figure}

Matrix multiplication is a widely researched topic and there has been a significant amount of research towards optimizing dense matrix-matrix multiplication (DGEMM) on the GPU. The BLAS~\cite{blas,blas2} library provides routines for basic vector and matrix operations, including matrix-matrix multiplication. Optimized libraries such as ATLAS~\cite{atlas} and Intel MKL~\cite{mkl} are also available for a variety of platforms. In addition, libraries such as MAGMA~\cite{magma} and NVIDIA cuBLAS~\cite{cublas} also offer optimized implementations of matrix-matrix multiplications that can leverage multi-core processors and GPUs. The approaches to optimizing dense matrix multiplication algorithm ~\cite{kirk2013,volkov2008,sahni2013} have been well researched and are utilized in the development of our algorithm as described in this section. 

Given matrices $A$ and $B$ of appropriate dimensions, the na\"ive approach to matrix multiplication ported to the GPU is shown in Algorithm~\ref{alg:naive_gpu}. A single GPU thread computes one output element of the matrix $C$. In order to compute a single output of the output matrix, each thread has to copy one row and one column of matrices $A$ and $B$ respectively from global memory, compute the overloaded inner product from Equation~\ref{eq:popc} and copy the result back to global memory.

\begin{algorithm}
	\KwData{blockIdx, blockDim, threadIdx - Block and thread identifiers defined by CUDA}
	\KwData{A, B - 2D Arrays of type 32-bit unsigned integer}
	\KwResult{Popcount as described in Section ~\ref{sec:algo}}
    i = blockIdx.y * blockDim.y + threadIdx.y \\
    j = blockIdx.x * blockDim.x + threadIdx.x \\
    \For {k = 0 to N-1}{
        TMP = TMP + POPCOUNT( AND( XOR( A[i,k], B[k,j] ), A[i,k] )\\
    }
    C[i,j] = TMP\\
    \vspace{.25cm}
  \caption{A na\"ive CUDA based implementation of the SNP comparison algorithm for $N_Q$ individuals and $N_R$ mixtures.}
  \label{alg:naive_gpu}
\end{algorithm}

\begin{description}[align=left]

\item[Tiling and Shared Memory usage] The na\"ive approach to matrix multiplication described earlier is bandwidth bound. The number of global memory transfers can be reduced by improving data locality through tiling and the use of shared memory. The tiling approach involves computing the output for a small block at a time and re-using the data already fetched from global memory. The GPU threads load a block of data required to compute a sub-block $C_{sub}$ of the output matrix $C$ into shared memory. The required sub-matrices $A_{sub}$ and $B_{sub}$ are loaded into the shared memory of a given block of threads and are used for computing the output matrix $C_{sub}$. This approach is illustrated in Figure~\ref{fig:block_matmul}. In this paper, block sizes of 16, 32 and 64 were used depending on the number of SNPs in the data being analyzed.

\item[Compute optimization] In addition to the tiled approach, a second optimization technique proposed by Volkov~\cite{volkov2010} is to compute more elements of the output matrix $C_{sub}$ per thread. This allows the use of fewer threads leading to a greater use of registers and more computations being performed in parallel. In this paper we compute 16 output elements per thread.
We also employ loop unrolling to unroll inner loops in the CUDA kernel that are not unrolled by the NVIDIA compiler by default.

\item[Memory access coalescing]
Two dimensional arrays in C/C++ are stored in row-major format. As a result, the memory accesses to the matrix $A$ by threads in a block are coalesced; i.e., threads in a wrap access successive memory locations in the GPU global memory. By coalescing memory accesses, the number of clock cycles required to fetch data from global memory to shared memory can be minimized. If memory accesses are not coalesced, the global memory access is effectively serialized. By transposing matrix $B$ in memory before transferring it to the GPU device, memory access to $B$ can also be coalesced. The memory layout of matrices $A$ and $B$ is adjusted appropriately while reading in the data from input files.

\end{description}

\subsection{Comparing Large Numbers of DNA Mixtures}
GPUs have a limited amount of RAM. The experiments described in this paper were conducted using a NVIDIA Tesla K80 GPU with 12GB of RAM. This limits the size of the matrices that are created in a kernel. For example, comparing 1,000,000 known DNA profiles with 2048 unknown profiles, each of length $N_W$, represented using 32-bit unsigned integers, generates a result matrix $C$ of size 2048 x 1,000,000 that requires 65GB of memory. To compare large numbers of DNA mixtures, we break up the computation into a series of smaller comparisons.

Moving data between the GPU memory space and the CPU memory space can be a significant bottleneck in GPU computing. One technique for hiding latency in data transfers between the GPU and CPU is to overlap compute with the data transfers. However, in our case, the entire memory available on the GPU is used for storing the inputs and the results of the DNA comparison algorithm in order to minimize the number of GPU kernel launches and the number of data transfers between the CPU and GPU. As a result, it is not possible to overlap the compute with data transfers. Typically the number of unknown DNA profiles is significantly smaller than the number of known reference profiles. In this case, we transfer all the query profiles and a block of known reference profiles to the GPU, followed by a GPU kernel launch to perform the comparisons. The next batch of known profiles to compare against is transferred to the GPU at the same time that the results from the previous batch are copied back to the CPU.

\begin{figure}[ht]
  \centering
  \subfloat[Number of unknown samples = 512]{
    \includegraphics[width=20pc]{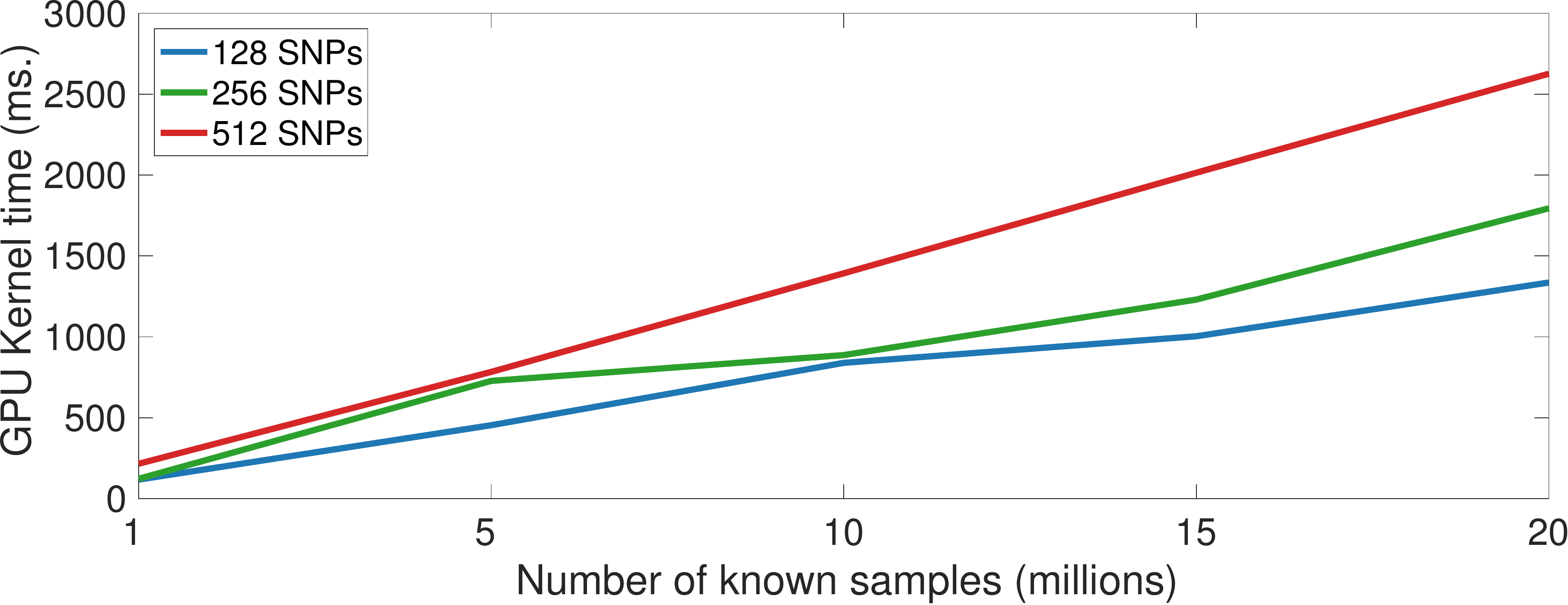}
    \label{fig:timing_w16} } \\
  \subfloat[Number of unknown samples : 1024]{
    \includegraphics[width=20pc]{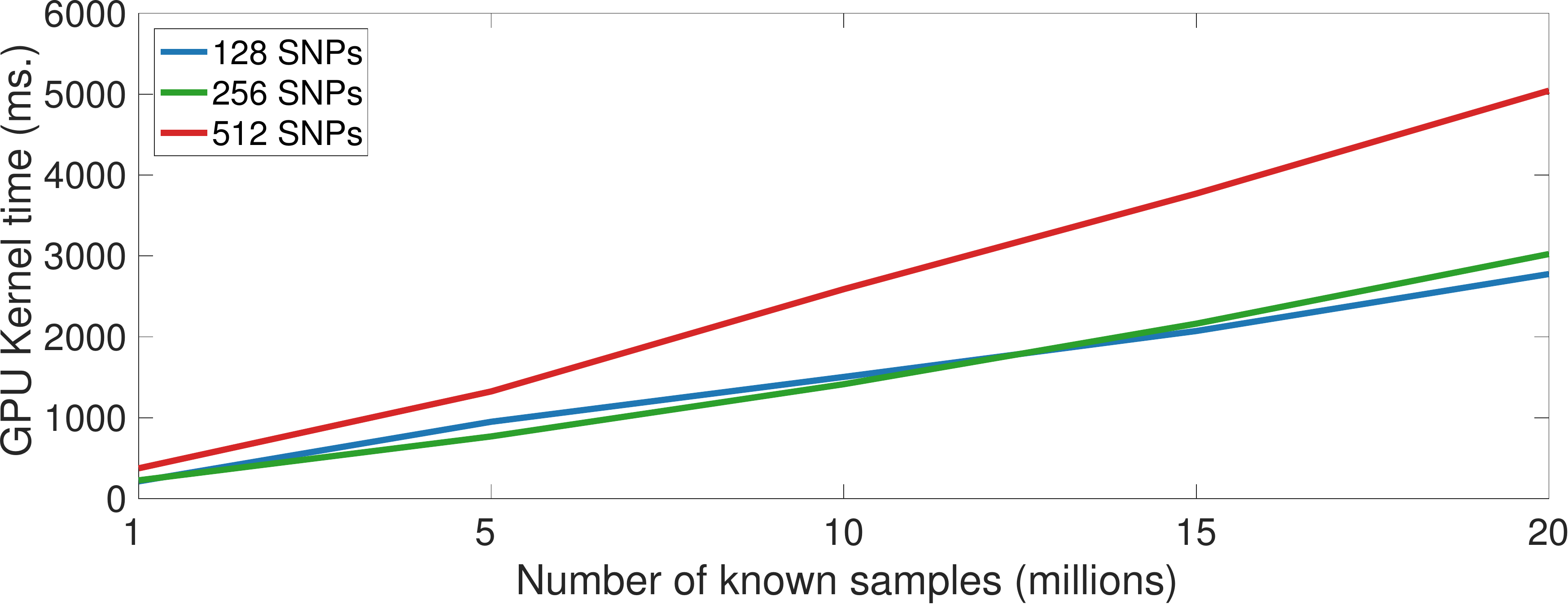}
    \label{fig:timing_w32} } \\
  \subfloat[Number of unknown samples : 2048]{
    \includegraphics[width=20pc]{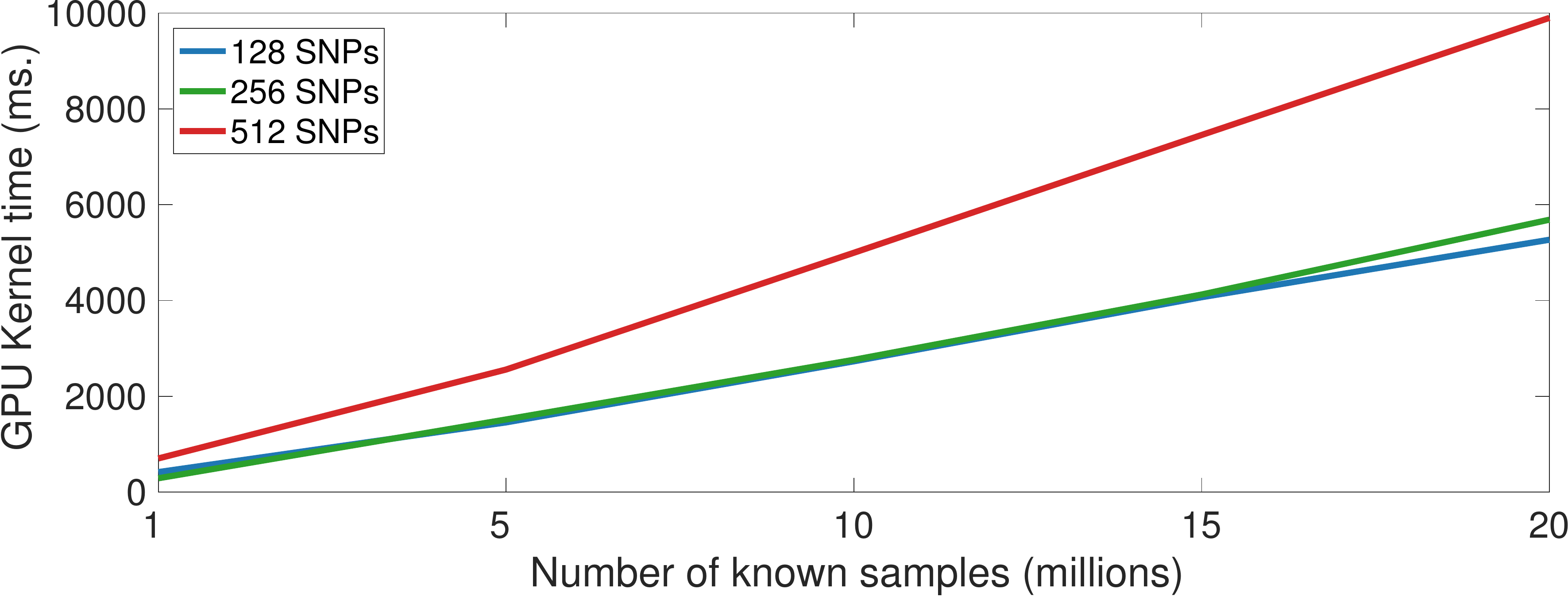}
    \label{fig:timing_w64} }
  \caption{GPU Kernel time for comparing unknown DNA profiles against 1M, 5M, 10M, 15M and 20M target profiles using SNPs of length 128, 256 and 512.}
  \label{fig:a}
\end{figure}

\section{Results}
\label{sec:results}
To test the performance of the proposed algorithm for comparing DNA mixtures, we compared 512, 1024 and 2048 unknown DNA profiles against 1, 5, 10, 15 and 20 Million known profiles.  Because of the large mismatch the number of known and unknown profiles, all unknown profiles were transferred to the GPU along with a block of known profiles. Depending on the total number of comparisons to be performed, the number of known reference profiles used in a given kernel launch was changed such that all memory on the GPU was utilized. This also helped minimize the number of data transfers between the CPU and GPU memory. As a result of nearly full utilization of GPU memory for each kernel launch, it was not possible to overlap data transfers and computation. Experiments were also perfomed to measure the performance of using pinned and non-pinned memory in the GPU kernel.

\begin{figure}[hb]
  \centering
  \subfloat[Number of unknown samples : 512, SNP length : 128]{
    \includegraphics[width=16pc]{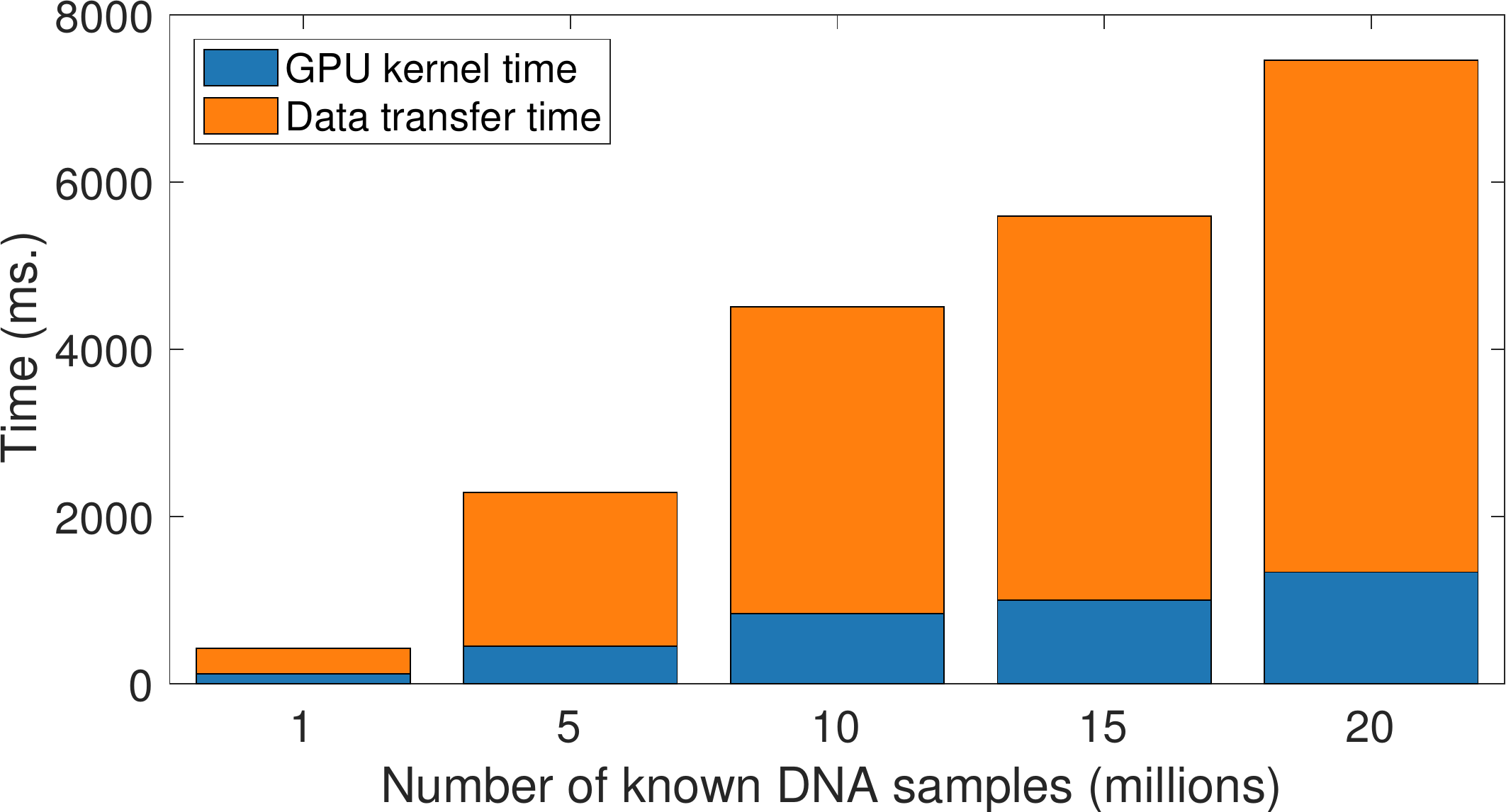}
    \label{memcpy_16} } \\
  \subfloat[Number of unknown samples : 2048, SNP length : 512]{
    \includegraphics[width=16pc]{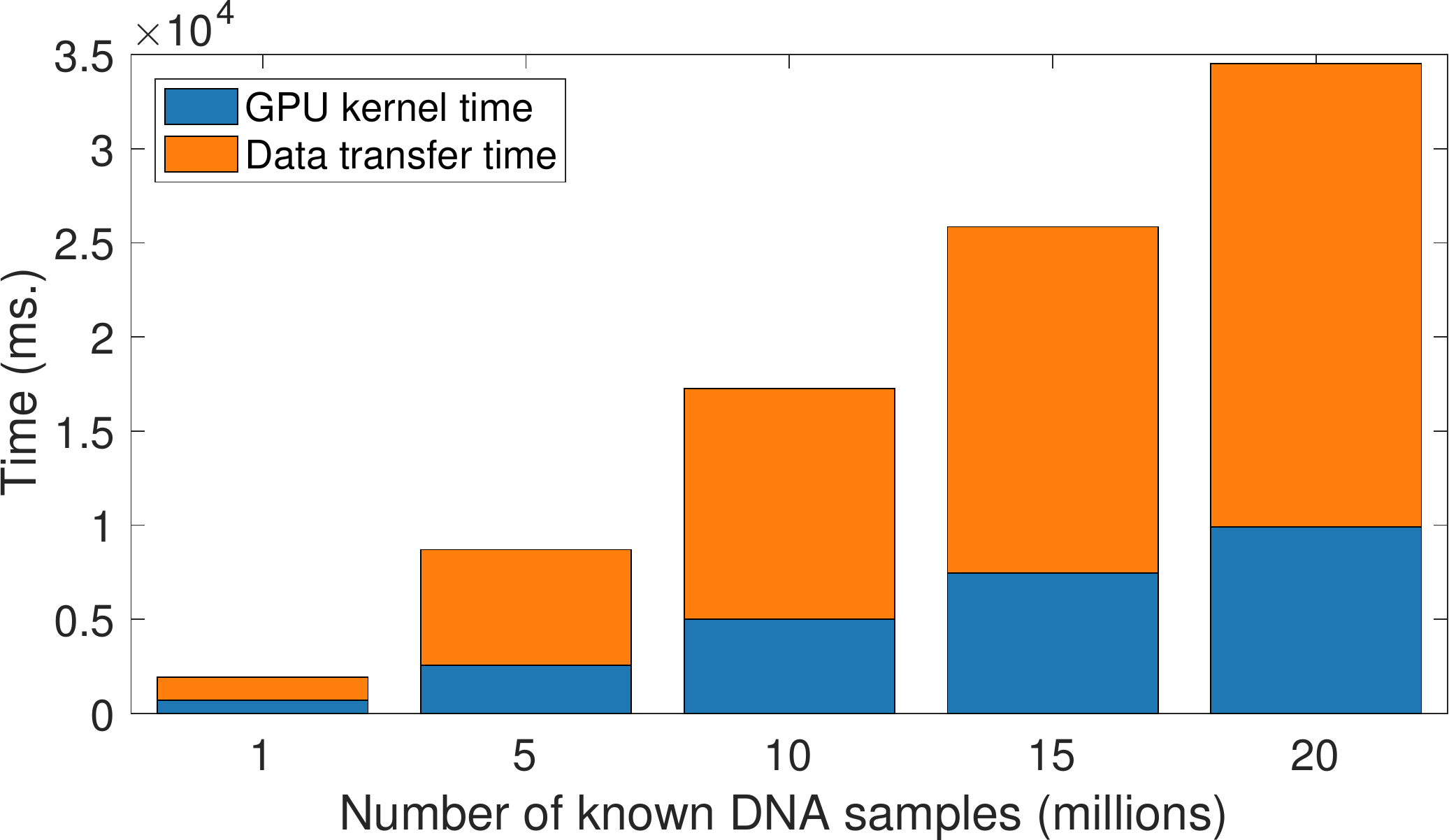}
    \label{memcpy_64} }
  \caption{Cummulative GPU kernel time and data transfer time for comparing known and unknown DNA profiles: As data size increases, the ratio of compute to data transfers improves.}
  \label{fig:memcpy_comparison}
\end{figure}

\begin{figure}[ht]
  \centering
  \includegraphics[width=20pc]{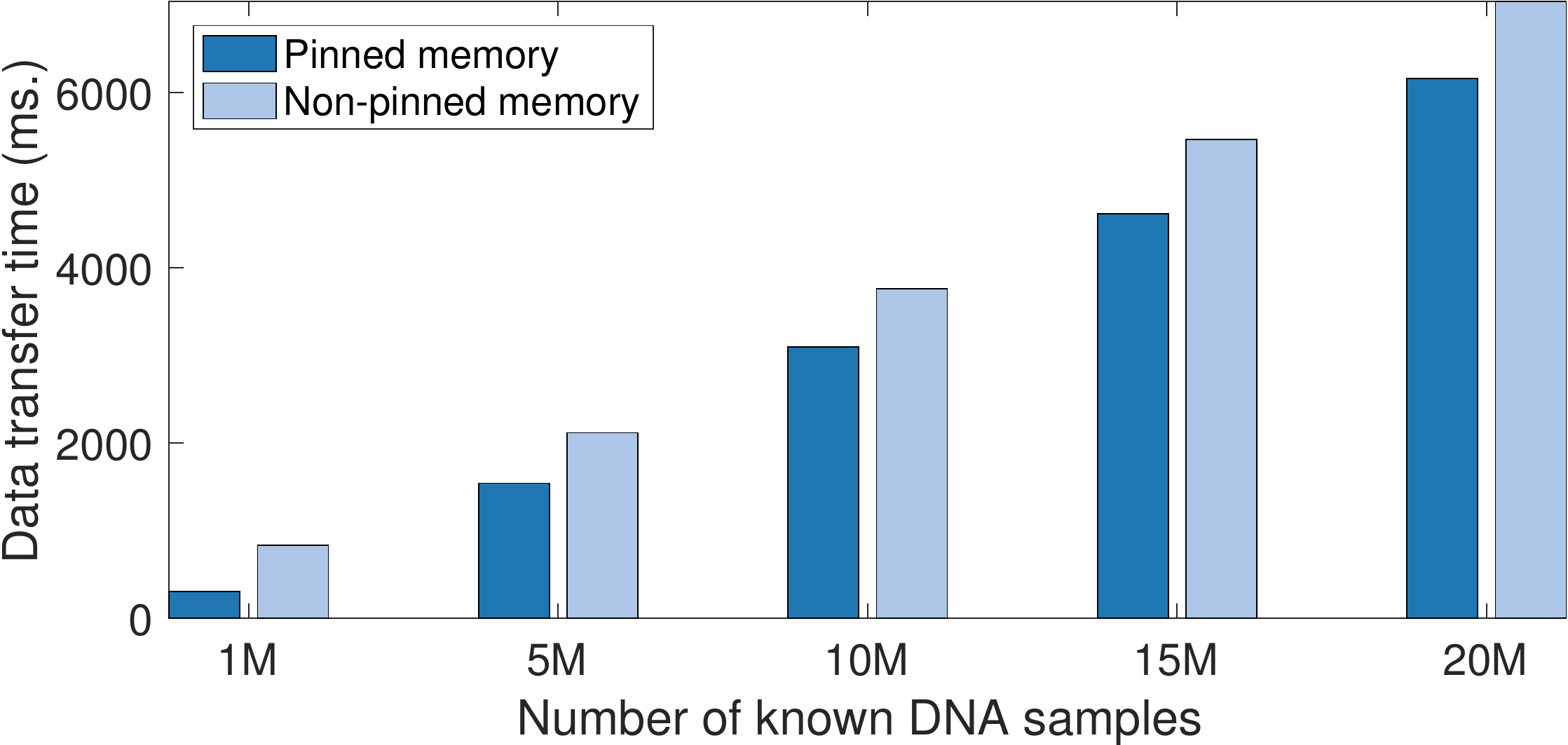}
  \caption{Comparison of data transfer time using pinned memory and non-pinned memory: Results shown for SNP length of 512.}
  \label{fig:timing_pin_nopin}
\end{figure}

Figures ~\ref{fig:timing_w16},~\ref{fig:timing_w32} and~\ref{fig:timing_w64} show the cumulative GPU kernel time for comparing DNA mixtures with 128, 256 and 512 SNPs respectively. While the total time spent in the GPU kernel is a function of the total number of comparisons between known and unknown DNA samples, the total time for the algorithm is dominated by the time required to transfer results back to the GPU. Transfer times for copying the known and unknown DNA samples to the GPU are a significantly smaller fraction of the total time spent in data transfers because of the relatively small amount of data being copied. Figure~\ref{fig:memcpy_comparison} shows the cummulative GPU kernel time and the total time spent in data transfers between the GPU and the CPU memory. As seen in this figure, the time spent in transferring data between the CPU and GPU tends to dominate. This time can be reduced by offloading additional computations to the GPU or performing additional reduction operation on the data in GPU memory. Additionally, the use of pinned memory can reduce the time it takes to copy results back to the CPU memory as shown in Figure ~\ref{fig:timing_pin_nopin}. Using pinned memory provides a consistently faster data transfer time as compared with the use of non-pinned memory but this comes at the cost of a small added overhead at the time that the memory is allocated for the first time.

\section{Summary} 
\label{sec:summary}
In this paper we discuss the formulation of DNA forensics as a dense linear algebra problem. A GPU based approach is used to speed up computations that involve comparing millions of known DNA profiles with a few thousand unknown profiles. Current approaches to DNA forensics employed by the forensics community require large computing systems and can take hours. By using GPUs and overloaded matrix multiplication as desribed in this paper, it is possible to reduce the compute time required to process large amounts of data. In this paper we use a single NVIDIA K80 for computations but this approach can be extended to use mulitple GPUs on the same system for a further reduction in compute times. Additionally, this implementation can also be run on laptops with NVIDIA hardware.

\section*{Acknowledgement} The authors would like to thank Adam Michaleas and Michael Jones for their support with NVIDIA hardware and software configuration. We would also like to thank David Martinez for his support.

\balance
\bibliographystyle{IEEEtran} \bibliography{IEEEabrv,references.bib}

\end{document}